\documentstyle[aps,prl,graphicx,floats,12pt]{revtex} 
\newif\ifoldrevtex \oldrevtextrue
\newif\iftwocolumns \twocolumnsfalse

\ifoldrevtex
 \newcommand{\affiliation}{\address}
 \newcommand{\email}[1]{\relax}
\fi

\renewcommand{\apj}[3]{Astrophys.\ J.\ {\bf #1}, #3 (#2)}
\renewcommand{\prl}[3]{Phys.\ Rev.\ Lett. {\bf #1}, #3 (#2)}
\renewcommand{\prd}[3]{Phys.\ Rev.\ {\bf D#1}, #3 (#2)}

\newcommand{\rhom}{\rho_M}

\newcommand{\rhocard}{\rho_{\rm Card}}
\newcommand{\rhoint}{\rho_{\rm internal}}

\begin{document}

\ifoldrevtex\iftwocolumns\twocolumn[\hsize\textwidth\columnwidth\hsize\csname
@twocolumnfalse\endcsname\fi\fi


\ifoldrevtex
\title{\vspace{-\baselineskip}\hfill\hbox{\footnotesize\rm CWRU-P17-02}\\ 
  \vspace{-0.6\baselineskip}\hfill\hbox{\footnotesize\rm NSF-ITP-02-173}\\
  An Accelerating Universe from Dark Matter Interactions with Negative
  Pressure} \else \preprint{CWRU-P17-02} \preprint{NSF-ITP-02-173} \title{An
  Accelerating Universe from Dark Matter Interactions with Negative Pressure}
\fi

\author{Paolo Gondolo} 
\email{pxg26@po.cwru.edu} 
\affiliation{Case Western Reserve University, Department of Physics, 10900
  Euclid Ave, Cleveland, OH 44106-7079}

\author{Katherine Freese}
\email{ktfreese@umich.edu}
\affiliation{Michigan Center for Theoretical Physics, University of Michigan,
  Ann Arbor, MI 48109, USA}

\ifoldrevtex\maketitle\vspace{\baselineskip}\fi

\begin{abstract}
  As an explanation for the acceleration of the universe, we propose dark
  matter with self-interactions characterized by a negative pressure; there is
  no vacuum energy whatsoever in this Cardassian model.  These
  self-interactions may arise due to a long-range ``fifth force'' which grows
  with the distance between particles.  We use the ordinary Friedmann equation,
  $H^2 = 8 \pi G \rho/3$, and take the energy density to be the sum of two
  terms: $\rho = \rho_M + \rho_K$.  Here $\rho_M \equiv m n_M$ is the ordinary
  mass density and $\rho_K \equiv \rho_K(\rho_M)$ is a new interaction term
  which depends only on the matter density.  For example, in the original
  version of the Cardassian model, $\rho_K = b \rho_M^n$ with $n<2/3$; other
  examples are studied as well.  We use the ordinary four-dimensional
  Einstein's equations $G_{\mu\nu} = 8 \pi G T_{\mu\nu} $ and assume a perfect
  fluid form for the energy momentum tensor.  Given this ansatz, we can compute
  the accompanying pressure, and use the conservation of the energy-momentum
  tensor to obtain modified forms of the Euler's equation, Poisson equation,
  and continuity equation.  With this fully relativistic description, one will
  then be able to compute growth of density perturbations, effects on the
  Cosmic Background Radiation, and other effects with an eye to observational
  tests of the model.
\end{abstract}


\ifoldrevtex\iftwocolumns\vskip2.0pc]\fi\else\maketitle\fi

\section{Introduction}

Observations of Type IA Supernovae \cite{SN1,SN2}, as well as
concordance with other observations, including the microwave
background \cite{boom} and galaxy power spectra \cite{2df}, indicate
that the universe is flat and accelerating.  Many authors have
explored possible explanations for the acceleration: a cosmological
constant, a decaying vacuum energy \cite{fafm,frieman}, quintessence
\cite{stein,caldwell,huey}, and gravitational leakage into extra
dimensions \cite{ddg}.

Freese and Lewis \cite{freeselewis} proposed Cardassian
expansion as an explanation for acceleration which invokes no vacuum
energy whatsoever.  In this model the universe is flat and
accelerating, and yet consists only of matter and radiation.  They
proposed additional terms in the Friedmann equation, which becomes
$H^2 = g(\rhom)$, where $\rhom$ contains only matter and radiation (no
vacuum), $H=\dot a/a$ is the Hubble constant (as a function of time),
$G=1/m_{pl}^2$ is Newton's universal gravitation constant, and $a$ is
the scale factor of the universe.  The function $g(\rhom)$ returns to
the usual $8\pi \rhom/(3m_{pl}^2)$ during the early history of the
universe, but takes a different form that drives an accelerated
expansion after a redshift $z \sim 1$.  Such modifications to the
Friedmann equation may arise, e.g., as a consequence of our observable
universe living as a 3-dimensional brane in a higher dimensional
universe \cite{cf}.

In this paper we propose an alternative origin for the same behavior.
This origin is a ``fifth force'' in ordinary four spacetime
dimensions.  We use the standard Friedmann equation
\begin{equation}
\label{standard}
H^2 = 8 \pi G \rho/3 ,
\end{equation}
but allow the dark matter to have self-interactions that contribute a
negative pressure: these self-interactions drive accelerated
expansion, and may arise due to a long-range ``fifth force.'' We
speculate on a form of the force between particles that may be
responsible for such an interaction: a confining force.

We take the energy density $\rho$ on the right hand side of
Eq.(\ref{standard}) to be the sum of two terms: the ordinary
contributions from matter and radiation plus a new interaction term.
During the matter dominated era, we take
\begin{equation}
\label{sum}
\rho = \rho_M + \rho_K ,
\end{equation}
where $\rho_M = m n_M$ (mass $m$ times number density $n_M$ of matter)
is the ordinary mass density and $\rho_K$ is a new interaction term
which is a function only of the mass density, $\rho_K \equiv
\rho_K(\rho_M)$.  We take the mass density to scale as usual with the
redshift,
\begin{equation}
\label{eq:matter}
\rhom = \rho_{M,0} (a/a_0)^{-3}
\end{equation}
where subscript $0$ refers to today.

For example, consider the original Cardassian model 
proposed in \cite{freeselewis}:
\begin{equation}
\label{eq:new}
H^2 = {8\pi\over 3 m_{pl}^2} \rhom + B \rhom^n \,\,\,\,\,\, {\rm with}
\,\,\,\,\,\, n<2/3 ,
\end{equation}
or, equivalently,
\begin{equation}
  \label{eq:friedcard}
  H^2 = \frac{ 8 \pi G}{3} \, \rhom \left[ 1 + 
    \left( \frac{\rhocard}{\rhom} \right)^{\!\!1-n\,} \right] ,
\end{equation}
where $\rhocard$ is the matter density at which the two
terms are equal.
The second term in square brackets is negligible
initially (when $ \rhom \gg \rhocard$), and only comes to dominate at
redshift $z \sim 1$ (once $\rhom \sim \rhocard$).  Once it dominates,
it causes the universe to accelerate: the scale factor grows as
$a \sim t^{2/3n}$ with $n<2/3$ so that $\ddot a > 0$.

There are two possible interpretations of Eq.(\ref{eq:friedcard}).
Previously, \cite{freeselewis} interpreted it as a modified Friedmann
equation which arose from the physics of extra dimensions \cite{cf}.
Here, and in the longer paper \cite{gondolo}, we treat this equation
as an ordinary Friedmann equation in which the second term describes
self-interactions of the dark matter 
particles\rlap,\footnote{This interpretation may have a four-dimensional
origin or serve as an effective description of higher dimensional physics.}
with
\begin{equation}
\rho_K = \frac{3
m_{pl}^2}{8\pi} B \rhom^n.
\end{equation} 

In a `generalized Cardassian model' other functions $g(\rhom)$ of the
matter density on the right hand side of the Friedmann
equation can also drive an accelerated expansion in the recent past of
the universe without affecting its early history \cite{freese,gondolo}.
Several of these alternative functions will be discussed below [see
Eqs.(\ref{eq:polytropic}) and (\ref{eq:modpoly})].

In the model we discuss here, we use the ordinary four-dimensional Einstein's
equations
\begin{equation}
G_{\mu\nu} = 8 \pi G T_{\mu\nu} .
\end{equation}
We take the energy momentum tensor to be made only
of matter and radiation, with the perfect fluid form
\begin{equation}
  T^{\mu\nu} = p g^{\mu\nu} + (p+\rho) u^{\mu} u^{\nu} ,
\end{equation}
where $\rho$ is the {\it total} energy density given in
Eq.(\ref{sum}), $p = p_M + p_K$ is the accompanying pressure, and
$u^\mu$ is the fluid four-velocity.  The Cardassian contribution to
the pressure can be computed in our class of models and turns out to
be a {\it negative pressure}, $p_K<0$.  This negative pressure is
responsible for the universe's acceleration.  We can find models with
any negative equation of state $w_K = p_K/\rho_K <0$, both constant
and time-dependent\footnote{We remark that in
our class of models, $w_K<-1$ is {\it not} connected to a violation of
the weak energy condition. Since $\rho_K$ and $p_K$ are not
independent of $\rho_M$, the weak energy condition is a condition
on the {\it total} energy-momentum tensor:
$(p_M+p_K)/(\rho_M+\rho_K) \ge -1$.}, including
$w_K<-1$; we also find models in which the pressure depends inversely
on the energy density, $p \sim - 1/\rho^r$.

We use the conservation of the energy-momentum tensor to obtain
modified forms of the Euler's equation, Poisson equation, and
continuity equation.  With this fully relativistic description, one
can then compute growth of density perturbations, effects on Cosmic
Background Radiation and cluster abundance, and other effects with an
eye to observational tests of the model.  An initial study of density
perturbations, performed in the Newtonian limit, can be found in our
longer paper \cite{gondolo}.

The Cardassian model also has the attractive feature that matter alone
is sufficient to provide a flat geometry.  The geometry of
Eq.(\ref{standard}) is flat, as required by observations of the
microwave background \cite{boom}.  The numerical value of the critical
mass density for which the universe is flat can be modified; we take
the value of the critical mass density to be 0.3 of the usual value.
Hence the matter density can have exactly this new critical value and
satisfy all the observational constraints, e.g., from the baryon
cluster fraction and the galaxy power spectrum.  Throughout this
paper, we will assume that there is {\it no vacuum energy} at all. We
do not solve the cosmological constant problem; we simply set it to
zero.

We speculate on the possible origin of an interaction energy with
negative pressure in Sect.~\ref{sec:confine}, present a general fluid
formulation in Sect.~\ref{sec:basics}, and then give specific examples
in Sect.~\ref{sec:examples}. We conclude in Section V.

\section{Origin of interaction energy with negative pressure}
\label{sec:confine}

Here we speculate on a possible origin for an interaction energy with
a negative pressure.  Dark matter particles may be subject to a new
interparticle force which is long-range and confining. This force may
be of gravitational origin or maybe a fifth force. 
It is the confinement property
that gives rise to a negative pressure (see \cite{gondolo}).  That a
confining force can give rise to an effective negative pressure is
well-known in particle physics, where the MIT bag model is just such
an effective description of quark confinement.

To be somewhat quantitative, let us consider a simple example of a
power law interparticle potential
\begin{equation}
  U_{ij} = A r_{ij}^\alpha ,
\end{equation}
where $\alpha>0$, $r_{ij}$ is the distance between particles and $A$
is a normalization constant. The corresponding force is a power law, $
F(r) \propto r^{\alpha-1} $.  The total new interaction energy of a
system of $N$ particles occupying a volume of radius $R$ will be $
U_{\rm new} \simeq A N^2 R^\alpha$, to within a numerical factor of
order 1 dependent on the geometry. The total gravitational potential
energy of the same system is, also within a factor of order unity,
 $ U_{\rm grav} \simeq \frac{ G M^2 }{R} ,$
where $ M$ is the total mass of the system. To play a cosmological
role at the present time, the new energy must be of the same order of
the gravitational energy when $R \simeq R_H$, the current size of the
horizon. Imposing that $U_{\rm new} \simeq U_{\rm grav}$ at $ R \simeq
R_H$ gives us the normalization
 $ A = \frac{G m^2}{R_H^{\alpha+1}} ,$
where $m=M/N$ is the mass of a single particle.  The magnitude of the
new force on galactic scales is negligible compared to the
gravitational force.  This Newtonian formulation must be
modified at large distances because of the finite speed of light and
issues of causality.  Notice that the Cardassian index $n$ in the
$\rho^n$ model is connected to the exponent $\alpha$ in the confining
force law through $\alpha = 3(1-n)$.

\section{Basic equations}
\label{sec:basics}

\subsection{Perfect Fluid}

As discussed in the Introduction, we use the ordinary four-dimensional
Einstein's equations $ G_{\mu\nu} = 8 \pi G T_{\mu\nu} $ where the
energy-momentum tensor is made only of matter and radiation and has
the perfect fluid form, $ T^{\mu\nu} = p g^{\mu\nu} + (p+\rho) u^{\mu}
u^{\nu} $.  Here the {\it total} energy density $\rho$ for matter includes
not only the mass density $\rhom$ (mass times number density) but also
any interactions as in Eq.(\ref{sum}).

\subsection{Conservation laws and Evaluation of Pressure}

The Bianchi identities guarantee the conservation of energy and momentum,
\begin{equation}
  {T^{\mu\nu}}_{;\nu} = 0 .
\end{equation}
In a comoving frame, energy-momentum conservation gives the (fully
relativistic) energy conservation and Euler equations
\cite{hawking,lyth-stewart}
\begin{eqnarray}
  \dot \rho & = & - \, {u^{\mu}}_{;\mu} \, (\rho + p) ,
  \label{eq:energycons} \\ \dot u_\mu & = & - \frac{ h^\nu_\mu
  p_{;\nu} }{ \rho + p } , \label{eq:Euler}
\end{eqnarray}
where the dot denotes a derivative with respect to comoving time and
the tensor $h_{\mu\nu} = g_{\mu\nu} - u_\mu u_\nu$ projects onto
comoving hypersurfaces.
We impose in addition that mass (or equivalently particle number) is
conserved,
\begin{equation}
( \rhom u^\mu )_{;\mu} = 0 .  
\end{equation}
This will give us the usual dependence of the mass density on the
scale factor of the universe, Eq.(\ref{eq:matter}).

{}From Eqs.(\ref{standard}) and (\ref{eq:energycons}), or
equivalently from adiabaticity, we can find the pressure due to the new
interactions,
\begin{equation}
  \label{eq:p-epsilon}
  p_{K} = \rhom \left( \frac{ \partial \rho_{K}} 
    {\partial \rhom} \right)_{\!\!s\,} -  
  \rho_{K} .
\end{equation}

\subsection{Newtonian limit}

Now we obtain the basic equations in the Newtonian limit.
In Minkowski space, we write $u^\alpha = \gamma(1,\vec{v})$
and the metric $\eta^{\alpha\beta} = {\rm diag}(-1,1,1,1)$, where
$\gamma = 1/\sqrt{1-v^2}$ and $\vec{v}$ is the 3-dimensional fluid velocity. 
Then, $T_{\mu\nu} =
{\rm diag}(-\rho,p,p,p)$.
We consider a weak static field produced by a nonrelativistic mass density.
Poisson's equation becomes
\begin{equation}
  \label{eq:poisson}
  \nabla^2 \phi = 4 \pi G (\rho + 3 p ) . 
\end{equation}
{}From ${T^{\alpha\beta}}_{;\beta} =0$ 
we find the continuity and Euler's equations
\begin{equation}
\label{eq:cont}
{\partial{\rho} \over \partial{t}} + \vec{\nabla} \cdot [(\rho 
+ p ) \vec{v}] = 0,
\end{equation}
\begin{equation}
\label{eq:nonrel-euler}
{\partial{\vec{v}} \over \partial{t}} + (\vec{v} \cdot \vec{\nabla})
\vec{v} = -  {\vec{\nabla}p \over \rho+ p} - \vec{\nabla}\phi.
\end{equation}
Notice that we do not assume $p \ll \rho$ in the right hand side of
Eqs.~(\ref{eq:poisson}-\ref{eq:nonrel-euler}).

With the additional constraint of particle number conservation,
we have the ordinary continuity equation for matter,
\begin{equation}
\label{eq:numbercons}
{\partial \rho_M \over \partial t} + \vec{\nabla} \cdot (\rho_M
\vec{v}) = 0 .
\end{equation}
Hence our basic nonrelativistic equations are Eqs.(\ref{eq:poisson} -
\ref{eq:numbercons}).

\section{Three Examples of Cardassian Models}
\label{sec:examples}

\subsection{Original $\rho^n$ Cardassian Model}

In the original Cardassian model of Ref.~\cite{freeselewis},
\begin{equation}
\label{yikes}
  \rho_{K} = b \rhom^n =
\rhom \left( \frac{\rhocard}{\rhom} \right)^{\!\!1-n\,},
\end{equation}
with $n<2/3$ as in Eqs.(\ref{eq:new}) and (\ref{eq:friedcard}). 
The pressure associated with this model in the fluid approach follows
from Eq.~(\ref{eq:p-epsilon}) as
\begin{equation}
  \label{eq:Keos}
  p_{K} = -(1-n) \rho_{K}.
\end{equation}
This model therefore has a constant negative $ w_{K} = p_{K}/\rho_{K}
= -(1-n) $.  Any observational test of $\rho^n$ Cardassian 
that depends only on $a(t)$ thus has similar behavior to quintessence;
however, tests that rely, e.g. on modified Poisson's equations will
have different observational consequences.

One can now obtain the basic equations in the Newtonian limit for the
$\rho^n$ Cardassian model by substituting $\rho_K$ and $p_K$ of
Eqs.(\ref{yikes}) and (\ref{eq:Keos}) into Eqs.(\ref{eq:poisson} -
\ref{eq:numbercons}).  In particular, the modified Poisson's equation
becomes
\begin{equation}
\label{eq:poisson2}
\vec{\nabla}^2 \phi = 4 \pi G \left[\rho_M - (2-3n) \left({\rhocard
\over \rhom}\right)^{1-n} 
\rho_M\right] ,
\end{equation}

The modified Euler's equation of Eq.(\ref{eq:nonrel-euler}) 
gives rise to a new force
\begin{equation}
\left. \frac{d\vec{v}}{dt} \right|_{\rm new}
= -{\vec{\nabla}p_K \over \rho_M} = + n(1-n) \left({\rhocard
\over \rhom} \right)^{1-n}
{\vec{\nabla}\rhom \over \rho_M} .
\end{equation}
On galactic scales, 
this force destroys flat rotation curves (velocities tend to increase
as one goes out to large radii in an unacceptable way).  This problem
for rotation curves persists in a relativistic generalization of this
argument. The fluid $\rho^n$ Cardassian case must therefore be thought
of as an effective model that applies only on cosmological scales.
The examples in the following two subsections present alternatives to
address the issues on galactic
scales. 

\subsection{Polytropic Cardassian}

Another class of models has
\begin{equation}
\label{eq:polytropic}
  \rho = \rhoint + 
  \rhocard \left[ 1 + \left( \frac{\rhom}{\rhocard} 
    \right)^{\!\!q\,}\right]^{ \frac{1}{q} }
\end{equation}
with $q \ne 0$.  This model can be used on all scales (see below), but
it does not quite fit the criteria of ``Cardassian'' since it does
have a vacuum term.  At late times in the future of the universe, when
$\rhom \ll \rhocard$, this model becomes cosmological constant
dominated with $\Lambda = \rhocard$.  
Eq.(\ref{eq:polytropic}) is very similar to a model that was
derived earlier \cite{ddg} due to gravitational leakage into
extra dimensions.

Here, the pressure is
\begin{equation}
  p = p_M - \rhocard \left[ 1 + \left( \frac{\rhom}{\rhocard} 
    \right)^{\!\!q\,}\right]^{ \frac{1}{q}-1} .
\end{equation}
When the ordinary pressure
$p_M$ can be neglected, this model obeys a polytropic equation of state
\begin{equation}
  p = - \rhocard \left( \frac{\rhocard}{\rho} \right) ^ {q-1} ,
\end{equation}
with negative pressure and negative polytropic index $N=-1/q$. For $q
> 1$, the pressure varies inversely with the energy density (to the
power $q-1$).

We must
make sure that at the scales of galaxies and clusters the Cardassian pressure
can be neglected compared to the ordinary pressure.  At large matter densities,
the Cardassian pressure is $ |p_{K}| \simeq \rhom^{1-q} \rhocard^{q} $. In a
galaxy or cluster with velocity dispersion $\sigma$, the ordinary pressure is
$p_M \simeq \rhom \sigma^2$. We want $ | p_K | / p_M \simeq (\rhocard/\rhom) ^
{q}/\sigma^2 \ll 1$.  Taking $\sigma \simeq 300 $ km/s and assuming $p_K$ is
unimportant out to $\approx$ 100 kpc where $ \rhom \approx 10^2 \, \rhocard$,
this condition amounts to $ q \gtrsim 3$. 

The models that match the rotation curves  have pressure that
scales inversely with the energy density. The condition $q\gtrsim
3$ requires that $p
\propto -{1 \over \rho^r}$ where $r=q-1 \gtrsim 2$.  This model is
similar to the Chaplygin gas \cite{chaplygin} which has $r\leq 1$;
note that we differ from the Chaplygin gas in the value of the
exponent that we find to be required.

Fits to supernova data \cite{wang} show that $ q \lesssim 2$. 
Hence we find difficulties in making polytropic cardassian models, or
generalized Chaplygin gas models, work at both cosmological and galactic
scales.

\subsection{Modified polytropic Cardassian}

Let us return now to the original Cardassian proposal in which
there is no vacuum energy whatsoever.
A Cardassian model that can be used on all scales is
\begin{equation}
\label{eq:modpoly}
  \rho = \rhoint + \rhom \left[ 1 + \left(
  \frac{\rhocard}{\rhom} \right)^{\!\!q(1-n)\,}\right]^{
  \frac{1}{q} }.
\end{equation}
For $q=1$ this reduces to the original $\rho^n$ Cardassian model. 
The pressure follows as
\begin{equation}
  p = p_M - (1-n) \rhom \left[ 1 + \left(
    \frac{\rhocard}{\rhom} \right)^{\!\!q(1-n)\,}\right]^{
    \frac{1}{q}-1} \left( \frac{\rhocard}{\rhom}
    \right)^{\!\!q(1-n)\,} .
\end{equation}
This model is interesting because the two parameters $n$ and $q$ are
important on different scales.  The parameter $n$ sets the current
value of $w \simeq - (1-n)$, and so can be chosen to fit the supernova
data, while the parameter $q$ governs the suppression of the
Cardassian pressure at high densities, and can therefore be chosen not
to interfere with galactic rotation curves and cluster dynamics. On
galactic scales, the pressure again depends inversely on the
energy density, but on large scales of the universe it depends linearly.
Concrete comparisons with data will be presented in \cite{wang}.

\section{Conclusions}

An interpretation of Cardassian expansion as an interacting dark
matter fluid with negative pressure is developed. The Cardassian term
on the right hand side of the Friedmann equation (and of Einstein's
equations) is interpreted as an interaction term. The total energy
density contains not only the matter density (mass times number
density) but also interaction terms. These interaction terms give rise
to an effective ngative pressure which drives cosmological
acceleration.
These interactions may be due to interacting dark matter, e.g. with a
long-range confining force or a fifth force between
particles. Alternatively, such interactions may be an effective
description of higher dimensional physics.

A fully relativistic fluid model of Cardassian expansion has been
developed, in which energy, momentum, and particle number are
conserved, and the modified Poisson's and Euler's equations have been derived.
In \cite{gondolo}, a
preliminary study of density fluctuations in the early universe has
also been presented.
There we developed a Newtonian theory
of perturbations, but discovered curious gauge ambiguities
in the relativistic theory that must be resolved in a future study.

One of our goals is to allow predictions of various
observables that will serve as tests of the model.  The Cardassian
model will have unique predictions, particularly due to the modified
Poisson's equation. For example, one can now calculate the effect on
the Integrated Sachs Wolfe component in the Cosmic Microwave
Background, as well as the effect on cluster abundances at different
redshifts.  These predictions can then be tested against measurements
of these quantities.  Comparison with existing and upcoming supernova
data is being studied in another paper \cite{wang}.

\acknowledgments

We thank T. Baltz, J. Frieman, and Y. Wang for useful conversations.  K.F.\ 
acknowledges support from the DOE via the University of Michigan, and thanks
the Aspen Center for Physics for hospitality. P.G.\ thanks the Michigan Center
for Theoretical Physics for hospitality.  This research was supported in part
by the National Science Foundation under Grant No. PHY99-07949 at the Kavli
Institute for Theoretical Physics, Santa Barbara, CA.



\begin{thebibliography}{99}
  
\bibitem{SN1} S.~Perlmutter {\it et al.}  [Supernova Cosmology Project
  Collaboration], Astrophys.\ J.\ {\bf 517}, 565 (1999) [astro-ph/9812133].
  
\bibitem{SN2} A.~G.~Riess {\it et al.}  [Supernova Search Team
  Collaboration], Astron.\ J.\ {\bf 116}, 1009 (1998)
  [astro-ph/9805201].
  
\bibitem{boom} C.B.~Netterfield {\it et al}, astro-ph/0104460; R.~Stompor {\it
    et al}, astro-ph/1015062; N.W.~Halverson {\it et al}, astro-ph/0104489; C.
  Pryke {\it et al} \apj{568}{2002}{46}

\bibitem{2df} L.~Verde {\it et al.} [2dF Survey] MNRAS {\bf 335}, 432 (2002)

\bibitem{fafm} K. Freese, F.C. Adams, J.A. Frieman, and E. Mottola,
  Nucl.\ Phys.\ {\bf B287}, 797 (1987).
  
\bibitem{frieman} J. Frieman, C. Hill, A. Stebbins, and I. Waga, \prl{75}
  {1995}{2077}.

\bibitem{stein} L. Wang and P. Steinhardt, \apj{508}{1998}{483}.
  
\bibitem{caldwell} R. Caldwell, R. Dave, and P. Steinhardt,
  \prl{80}{1998}{1582}
  
\bibitem{huey} G. Huey, L. Wang, R. Dave, R. Caldwell, and P. Steinhardt,
  \prd{59}{1999}{063005}
  
\bibitem{ddg} C.~Deffayet, G.~Dvali, and G.~Gabadadze, Phys.\ Rev.\
  {\bf D65}, 044023 (2002).

\bibitem{freeselewis} K.~Freese and M.~Lewis, Phys.\ Lett.\ {\bf B540}, 1
  (2002) [astro-ph/0201229].

\bibitem{cf} D. Chung and K. Freese, Phys. Rev. D {\bf 62}, 063513 (2000).

\bibitem{gondolo} P. Gondolo and K. Freese,  hep-ph/0209322
  
\bibitem{freese} K.~Freese, hep-ph/0208264.

\bibitem{hawking} S.~Hawking, Ap. \ J. \ {\bf 145}, 544 (1966).

\bibitem{lyth-stewart} D.~Lyth and E.~Stewart, 
Ap. \ J. \ {\bf 361}, 343 (1990).

\bibitem{wang} Y.~Wang, K.~Freese, J.~Frieman, P.~Gondolo, and M.~Lewis,
in preparation.

\bibitem{chaplygin} A. Kamenshchik, U. Moschella, V. Pasquier, Phys.\ Lett.\
  {\bf B551}, 265 (2001); M.C. Bento, O. Bertolami, and A.A. Sen,
astro-ph/0210468; N. Bilie, G.B. Tupper, and R.D. Viollier,
astro-ph/0207423; V. Gorini, A. Kamenshchik, and U. Moschella,
astro-ph/0209395.

\end{thebibliography}
\end{document}